\documentclass[preprints,article,submit,moreauthors,pdftex]{Definitions/mdpi} 
\renewcommand{\linenumbers}{}

\usepackage[sort&compress]{natbib}
\usepackage{braket,mleftright}
\usepackage{ dsfont }
\usepackage{graphicx}
\usepackage[justification=centering]{subfig}

\pdfoutput=1

\Title{Noise-Assisted Discord-Like Correlations in Light-Harvesting Photosynthetic Complexes}

\TitleCitation{Noise-Assisted Discord-Like Correlations in Light-Harvesting Photosynthetic Complexes}

\Author{Pablo Reséndiz-Vázquez $^{1}$, Ricardo Rom\'an-Ancheyta $^{2}$ and Roberto de J. León-Montiel $^{1,}$*}

\AuthorNames{Pablo Reséndiz-Vázquez, Ricardo Rom\'an-Ancheyta and Roberto de J. León-Montiel}

\AuthorCitation{Reséndiz-Vázquez, P.; Rom\'an-Ancheyta, R.; \mbox{de J. León-Montiel, R.}}

\address{%
$^{1}$ \quad Instituto de Ciencias Nucleares, Universidad Nacional Autónoma de México, Apartado Postal 70-543,  \mbox{Ciudad de M\'exico 04510}, Mexico; pablorv@ciencias.unam.mx \\
$^{2}$ \quad  Instituto Nacional de Astrof\'isica, \'Optica y Electr\'onica, Calle Luis Enrique Erro 1, Sta. Ma.  Tonantzintla, Puebla CP 72840, Mexico; ancheyta6@gmail.com}

\corres{Correspondence:  roberto.leon@nucleares.unam.mx}

\abstract{Transport phenomena in photosynthetic systems have attracted a great deal of attention due to their potential role in devising novel photovoltaic materials. In particular, energy transport in light-harvesting complexes is considered quite efficient due to the balance between coherent quantum evolution and decoherence, a phenomenon coined Environment-Assisted Quantum Transport (ENAQT). Although this effect has been extensively studied, its behavior is typically described in terms of the decoherence's strength, namely weak, moderate or strong. Here, we study the ENAQT in terms of  quantum correlations that go beyond entanglement. Using a subsystem of the Fenna--Matthews--Olson complex, we find that discord-like correlations maximize when the subsystem's transport efficiency increases, while the entanglement between sites vanishes. Our results suggest that quantum discord is a manifestation of the ENAQT and highlight the importance of beyond-entanglement correlations in photosynthetic energy transport processes.}


\begin{document}

%

\section{Introduction}

Transport phenomena in nanostructured materials \cite{Stegmann2016,Veldhorst2012,Beenakker1991,Pablo2020} and biomolecules~\cite{Capasso, Plenio1, Zhang, Luis2020, Rob1} have been the main subject of interest in several investigations in the last two decades. Of particular importance is the study of energy transport and energy conversion, in photosynthetic complexes, from a quantum mechanical perspective~\cite{Mccree1971,Sension2007}. Remarkably, in some biological systems, such as green sulfur bacteria, where the corresponding energy transport efficiency is exceptionally high, there is experimental evidence of unexpected long-time quantum coherences~\cite{Engel2007,Calhoun2009,Lee2007}. \textcolor{black}{These observations have led to the proposal of different mechanisms through which excitonic energy transport may be enhanced. One of these is the so-called environment-assisted quantum transport or ENAQT, an effect that arises from the balance between coherent quantum evolution of a photosynthetic system and environmentally induced decoherence \cite{Alan,Plenio2}. More recently, the role of vibrations in efficient transport of photosynthetic energy has been highlighted, giving rise to the so-called vibrationally assisted energy transfer effect \cite{Dubi_2018,li2021}.}


On the other hand, quantum computation promises information processing with tremendous efficiencies that only quantum devices could handle~\cite{Huang2020}. The conventional view is that the primary resource for obtaining this enhancement is entanglement. Unfortunately, the creation and experimental manipulation of entangled states remain a technological challenge, as it requires extreme isolation from the surrounding environment. Hence, the search for quantum protocols that allow for significant enhancements in the efficiency of several quantum processes involving non-entangled states is of great interest~\cite{Merali2011}. In this sense, quantum discord is a valuable quantum information resource that could contribute to specific quantum processes with high efficiencies.

\textcolor{black}{The presence of genuinely quantum phenomena in complex biological processes, such as olfaction \cite{brooks}, magnetodetection \cite{Kominis2020}, and photosynthesis \cite{ball}, is still under debate.} In particular, the possibility that quantum effects may play a role in the light-harvesting process of bacteria and algae~\cite{Sarovar2010, Ishizaki2010, Whaley2011, Fassioli2010, Engel2007} has received several criticisms~\cite{Datta2005, Leon2013}. Nevertheless, the study of these effects underlying photosynthetic complexes and their technological implementation could pave the way to bio-inspired materials operating at ambiance temperatures with very high light transport and harvest efficiency.


In this short report, we study the excitonic transfer in a three-site subsystem of the Fenna--Matthews--Olson (FMO) complex. We show how quantum discord-like correlations maximize during the noise-assisted energy transport process in the single-excitation regime. Due to the dephasing noise, the off-diagonal elements of the density matrix washout. Consequently, in this regime, the bipartite entanglement between sites vanishes. Our results illustrate the emergence of correlations that go beyond-entanglement in open photosynthetic systems that may be present in bio-inspired materials, operating as efficient solar cells.



\section{The Model}

To explore the relationship between beyond-entanglement correlations and ENAQT, we consider a simple subsystem of three chromophores (sites) of the FMO complex of \textit{Prosthecochloris aestuarii} \cite{fenna_matthews,olson}. Under weak light illumination, the subsystem's energetic component resembles a thigh-binding Hamiltonian of the form~\cite{Rob1, May}
\begin{equation}\label{eq:eq1}
H_{C}= \sum_{i=1}^{3} \varepsilon_{i} \ket{i}\bra{i} + \sum_{i<j}^{3} V_{ij}\big(\ket{i}\bra{j}+\ket{j}\bra{i}\big),
\end{equation}
where $\varepsilon_{i}$ is the energy of the site $i$ and $V_{ij}$ the symmetric intermolecular coupling between sites $i$ and $j$. In the next section, we use explicit values of $\varepsilon_{i}$ and $V_{ij}$ that we take from the experimental data shown in Tables 2 and 4 of~\cite{Renger}. We denote the state where no excitations are present as $\ket{g}$, and $\ket{RC}$ as the state where the exciton (a bound state of an electron in a conduction band with a hole in the valence band) is transferred to the corresponding reaction center. We will use $|RC\rangle$ to compute the transport efficiency at different dephasing rates. Notice that, in our model, the states $\ket{g}$ and $\ket{RC}$ are not directly coupled to the subsystem's sites by the unitary evolution generated by $H_{C}$; instead, we will see how the open dynamics link them.

In general, describing the dynamics of photosynthetic systems interacting with their surrounding environment in full detail is a nontrivial task, mainly because this is non-Markovian~\cite{May,chen2011,mohseni2014}. However, we will use a Markovian model of the environment which, although oversimplified, includes the necessary physics to qualitative reproduce several experimental observations of energy-transport made in multi-chromophoric photosynthetic complexes~\cite{Haken1,Haken2,Kriete1,Moix2013}. 
In this regard, a Lindblad master equation for the density matrix $\rho$ of the FMO's subsystem can describe the influence of the environment upon this, and is given by~\cite{Petru}

\begin{equation}\label{eq:eq2}
\frac{\partial \rho}{\partial t} = -\frac{i}{\hbar}[H_{C},\rho] + \mathcal{L}_{deph}[\rho] + \mathcal{L}_{diss}[\rho] + \mathcal{L}_{RC}[\rho],
\end{equation}
\textls[-15]{where the first term on the right-hand side of this equation is just the unitary evolution generated by the Hamiltonian $H_C$. The following terms outline the open dynamics. For~instance, }
\begin{equation}\label{eq:eq3}
\mathcal{L}_{deph}[\rho]= \sum_{i\textcolor{black}{=1}}^{\textcolor{black}{3}} 2 \gamma_{i}\Big( \ket{i}\rho_{ii}\bra{i}-\frac{1}{2}\big{\{}\ket{i}\bra{i},\rho \big{\}} \Big)
\end{equation}

\noindent represents a pure dephasing process that makes any coherence (the off-diagonal elements $\rho$) be reduced exponentially at a rate $\gamma_{i}$; note that $\{.,.\}$ stands for the anticommutator. The third term in the Lindblad master Equation (\ref{eq:eq2}) is
\begin{equation}\label{eq:eq4}
\mathcal{L}_{diss}[\rho]= \sum_{i\textcolor{black}{=1}}^{\textcolor{black}{3}} 2 \Gamma_{i}\Big( \ket{g}\rho_{ii}\bra{g}-\frac{1}{2}\big{\{} \ket{i}\bra{i},\rho\big{\}} \Big)
\end{equation}
\noindent and models energy dissipation of the system to the environment. For example, an exciton could recombine in the site $i$ at rate $\Gamma_{i}$. However, the lifetime of the excitons in the FMO complex is usually larger ($\sim$ps) than the duration of their transport phenomena ($\sim$fs)~\cite{Luis2020,Rob1}. Thus, in several circumstances, $\Gamma_i$ can be effectively neglected.
%
The last term in Equation~(\ref{eq:eq2}) rules the irreversible transfer of excitations from a chromophore or site $\ket{k}$ to the reaction center $|RC\rangle$ at rate $\Gamma_{\texttt{RC}}$. Its explicit expression reads
\begin{equation}\label{eq:eq5}
\mathcal{L}_{RC}[\rho]= 2 \Gamma_{\texttt{RC}}\Big( \ket{RC}\braket{k|\rho|k}\bra{RC}-\frac{1}{2}\big{\{} \ket{k}\bra{k},\rho \big{\}} \Big).
\end{equation}

To quantify how efficient the transfer process of FMO's excitons to the reaction center is, we define the transport efficiency $\eta$ as the probability that the energy will arrive at the reaction center in a much longer time than the characteristic time of the FMO dynamics; 
this is given by~\cite{Rob1}
\begin{equation}\label{eq:eq6}
\eta \equiv \lim_{t \to \infty } \braket{RC|\rho|RC}.
\end{equation}

{In the next section, we use a single-excitation in one of the system's sites as our numerical simulations' initial condition. Previous studies of energy transport have also assumed such condition in photosynthetic light-harvesting complexes~\cite{Ishizaki2009,Alan2,Valleau2012,Setexciton,Saikin2017}.}

\subsection*{Quantum Correlations}

Quantum discord is a well-known quantifier of quantum correlations that go beyond-entanglement. For instance, one can have states with zero entanglement but non-zero discord~\cite{Henderson_2001,PRL_Zurek_2001}. In a bipartite system, one obtains the quantum discord by subtracting the classical correlations from the quantum mutual information; the latter measures the total (quantum and classical) correlations. Experimental and theoretical studies on quantum discord range from remote state preparation~\cite{Dakic2012}, correlated photonic systems~\cite{Dominguez2017}, ferromagnetic~\cite{Fedorova2019} and antiferromagnetic~\cite{Singh2015} materials. Remarkably, these works show that quantum discord may be a necessary resource for tasks to be realized with high efficiency.

To explore the presence of possible non-classical correlations in the system described above, we use a discord-like base measure known as Local Quantum Uncertainty (LQU). The LQU quantifies the minimum Wigner--Yanase skew information achievable on a single local measurement~\cite{Girolami}. Let $\rho_{_{\textcolor{black}{AB}}}$
be the state of a bipartite system such that, for this work, we assume one party to be formed by a single molecule (or site), while a subset of the remaining FMO's sites forms the other party, i.e., a qubit--qudit system that lives on a $\mathds{C}^{2} \otimes  \mathds{C}^{d}$ dimensional Hilbert space. For such a case, an analytical formula of the LQU with respect to subsystem A can be obtained and reads as~\cite{Girolami,Tian_2020}
\begin{equation}\label{eq:eq7}
\mathcal{U}_{A}(\rho_{_{\textcolor{black}{AB}}}) = 1 - \lambda_{max} \{W_{AB}\},
\end{equation}
where $\lambda_{max}$ denotes the maximum eigenvalue of the $3\times3$ symmetric matrix $W_{AB}$ whose elements are given by
\begin{equation}\label{eq:eq8}
(W_{AB})_{ij}\equiv {\rm Tr}\{\sqrt{\rho_{_{\textcolor{black}{AB}}}}(\sigma_{iA}\otimes \mathds{1}_{B})\sqrt{\rho_{_{\textcolor{black}{AB}}}}(\sigma_{jA}\otimes \mathds{1}_{B})\}.
\end{equation}
$\sigma_{iA}$ are the standard Pauli matrices of the qubit $A$ with $i,j=x,y,z$. \textcolor{black}{Note that $\rho_{_{AB}}$ stands for the density matrix of a system comprising all possible three-qubit states, whereas $\rho$, used in Equation~(\ref{eq:eq2}), stands for a density matrix involving  only the single-excitation-basis states.}
It is easy to show that in the single-excitation regime Equation (\ref{eq:eq7}) reduces to \textcolor{black}{(see Appendix \ref{app1} for details)}
\begin{equation}\label{eq:eq9}
\mathcal{U}_{S}(t) = 1-\sum_{l,m=1}^{3} \lambda_{l}^{1/2} \lambda_{m}^{1/2}\left| \braket{v_{l}|\sigma_{zA}\otimes\mathds{1}_{B}|v_{m}} \right| ^{2},
\end{equation}
where $\ket{v_{l}}$ is the eigenvector associated with the eigenvalue $\lambda_{l}$ of the density matrix $\rho$\textcolor{black}{, i.e., the state for the three-chromophore single-excitation system.} Subscript $S$ on $\mathcal{U}_S(t)$ denotes the single-excitation regime.

Finally, in order to monitor the changes in the LQU as ENAQT is activated, that is, as dephasing is introduced into the system, we examine the flux $\Phi_{LQU}$ of LQU through the photosynthetic system. This is defined as
\begin{equation}\label{eq:eq10}
\Phi_{LQU}^{\gamma}\equiv \mathcal{U}_{S}^{\gamma}(t \to \infty )- \mathcal{U}_{S}^{\gamma}(t = 0),
\end{equation}
where the superscript $\gamma$ stands for each of the increasingly larger dephasing rates considered in our numerical simulations.

\section{Results}

\begin{figure}[H]
\includegraphics[scale=0.65]{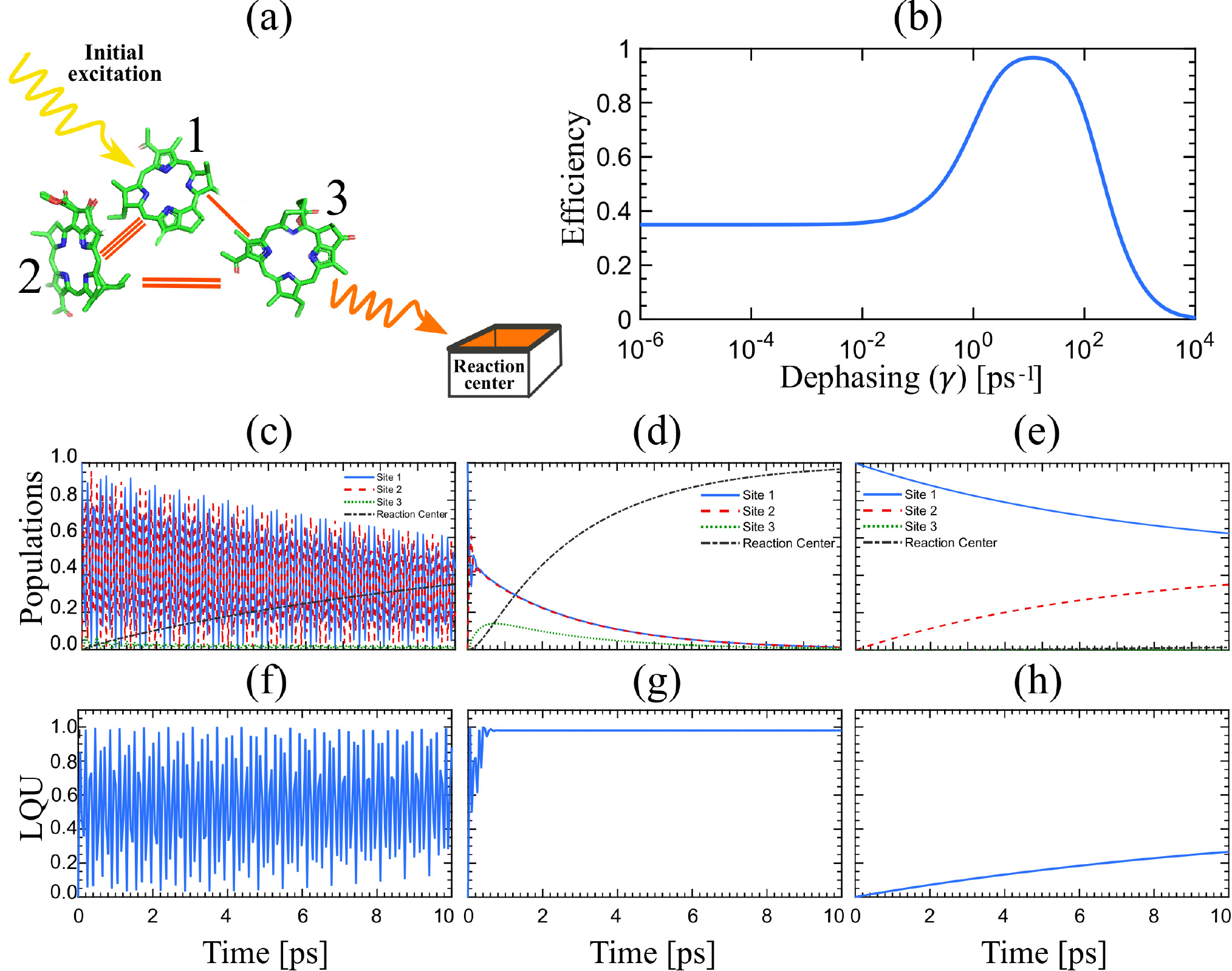}
\centering
\caption{(\textbf{a}) Schematic representation of a three-site system with energy sites \mbox{$\varepsilon_{1}= 215$ $\text{cm}^{-1}$}, \mbox{$\varepsilon_{2}= 220$~$\text{cm}^{-1}$}, $\varepsilon_{3}= 0$ $\text{cm}^{-1}$ and symmetric couplings $V_{12}=V_{21}= -104.1$ $\text{cm}^{-1}$, \mbox{$V_{13}=V_{31}= 5.1$~$\text{cm}^{-1}$} and $V_{23}=V_{32}= 32.6$ $\text{cm}^{-1}$. The initial condition is assumed to be localized in the first site, while the reaction center is connected through an irreversible loss-channel to the third site, with a transfer rate of $\Gamma_{\texttt{RC}}= 1.0$ $\text{ps}^{-1}$. (\textbf{b}) Energy transfer efficiency, $\eta$, as a function of the dephasing rate, $\gamma$. The evolution of the populations of the three sites is shown in figures (\textbf{c}--\textbf{e}) for $\gamma=10^{-6}$~$\text{ps}^{-1}$, $\gamma=12.07$~$\text{ps}^{-1}$ and $\gamma=10^{4}$ $\text{ps}^{-1}$, respectively. The time evolution of the Local Quantum Unicertainty (LQU) is shown in (\textbf{f}--\textbf{h}) for the same values of dephasing as for the populations. Note that conversion between units of $\text{cm}^{-1}$ and $\text{ps}^{-1}$ can be realized by making use of the equivalence $\hbar \sim 5.3$ $\text{cm}^{-1}~\text{ps}$.}\label{fig:figure1}
\end{figure}

In the following, we consider a subunit of the FMO complex of \textit{Prosthecochloris aestuarii}. Recall that this complex consists of seven coupled bacteriochlorophyll (BChl) \linebreak molecules~\cite{Alan,Plenio2,Rob1}; however, for the sake of simplicity, we work only with three of them, see Figure~\ref{fig:figure1}a for a schematic representation. The corresponding energies $\varepsilon_i$ and couplings $V_{ij}$ are taken from the three first chromophores of the FMO complex (following the original ordering of Fenna, Matthews and Olson \cite{fenna_matthews,olson}). \textcolor{black}{Note that this selection is not arbitrary; in the FMO complex, the site closest to the chlorosome antenna complex (the source of excitations) is site 1, whereas site 3 is the closest to the reaction center, implying that energy transfer from the FMO complex to the reaction center proceeds through that site~\cite{Renger}. In light of this information, we selected a subsystem that comprises these two important sites, whose energy transfer is mainly controlled by site 2 due to the weak coupling between sites 1 and 3.} \textcolor{black}{Interestingly, a similar trimeric chromophore system was used recently to explore the underlying physics of vibrationally-assisted energy transfer \cite{li2021}}. We follow previous authors~\cite{Plenio1,Rob1,Renger} that estimate $\Gamma_{\texttt{RC}}=1$~ps$^{-1}$. \textcolor{black}{As we already mentioned, the BChl 3 is the closest to the reaction center; therefore, we set $k=3$ in Equation~(\ref{eq:eq5}).} Furthermore, we take the dissipative and dephasing rates to be equal for all molecules, i.e.,  $\Gamma_{i} \equiv \Gamma= 5.0 \times 10^{-4}$~ps$^{-1}$~\cite{Plenio2,Rob1} and $\gamma_{i}\equiv \gamma$~\cite{Rob1}, respectively.  Finally, we consider the initial state of the system to be a localized one ~\cite{Plenio1,Plenio2,Ishizaki2009,Alan2,Alan3,Pelzer,Manzano}, that is,
\begin{equation}\label{eq:eq11}
\rho(0)= \ket{1}\bra{1}.
\end{equation}

As one might expect, the excitonic transport efficiency, quantified by $\eta$ in Equation~(\ref{eq:eq6}), strongly depends on the initial excitation conditions and system-bath dephasing interactions \cite{Rob1}. This can clearly be seen, in Figure~\ref{fig:figure1}b, as a significant enhancement of $\eta$ from $38\%$ to $97\%$.
%
%
Note that, in the regime where dephasing noise goes from $10^{-6}$~ps$^{-1}$ to $10^{-2}$~ps$^{-1}$, the environment is not strong enought to break the induced coherent localization. This localization is mainly caused by the intrinsic static disorder~\cite{Alan,Anderson} between the sites of Figure~\ref{fig:figure1}a. When the strength of the environmental noise increases, the efficiency $\eta$ reaches a maximum value of $97\%$. Such behavior coincides with previous studies on ENAQT~\cite{Alan, Rob1}, and can be understood as a result from a balance between the coherent quantum evolution and the incoherent dephasing process, which destroys any trace of localization and leads to an incoherently delocalized exciton state \cite{robertoLPL}. In this situation, the initial excitation is able to effectively reach the corresponding reaction center~\cite{Alan,Rob1}. Finally, when the dephasing rate is too strong, it acts as a constantly repeated measurement that inhibits the system's energy transport, i.e., it traps the exciton in its initial state~\cite{Alan, Plenio1,Plenio2,Rob1}.

One of the main goals of this work is to examine the emergence of beyond-en\-tan-glement correlations in terms of the environmental noise present in the FMO's subunit. As we next show, there is an interesting relation between the flux in LQU, $\Phi_{LQU}$, and the efficiency $\eta$ that we obtain by computing the LQU correlations by means of Equation~(\ref{eq:eq9}) on the subunit's density matrix $\rho(t)$. We evaluate the temporal evolution of the LQU at distinct regimes of dephasing for a $\mathds{C}^{2} \otimes  \mathds{C}^{4}$ Hilbert space given by the partition $\{ \{\ket{1} \}, \{ \ket{2},\ket{3}\}\}$. Note that we chose these two particular subsets so that the LQU provides information about the correlations between the initially excited state and the remaining sites.

We find that, in the coherent evolution regime, where $\gamma$ is smaller than $10^{-2}$~ps$^{-1}$, the LQU oscillates between zero-LQU and maximum-LQU, see Figure~\ref{fig:figure1}f. This behavior can be attributed to the strong interaction between sites 1 and 2, as depicted by the three lines that join them in Figure~\ref{fig:figure1}a. Hence, a typical coherent evolution regulates the system in such a regime (see Figure~\ref{fig:figure1}c). Interestingly, by increasing the dephasing rate $\gamma$, we observe a quick saturation of the LQU and the reaction center's population in the regime where ENAQT is present, see Figure~\ref{fig:figure1}g,d, respectively. Quite remarkable is that, due to the dephasing effect, the off-diagonal elements of the density matrix (coherences) are so small that, in this regime, any trace of entanglement is feeble~\cite{Sarovar2010}, \textcolor{black}{this suggests that the LQU, instead of entanglement, may be a resource for ENAQT in our system.} In Figure~\ref{fig:figure1}h, we see that when the excess of dephasing hinders the transport to the reaction center, the LQU correlations slowly increase during the time evolution. This happens due to the high probability of finding the exciton at either site 1 ($\approx$80\%) or site 2 (see Figure~\ref{fig:figure1}e).


\textcolor{black}{An intuitive understanding for the build-up of discord-like correlations is that the noise caused by dephasing allows the excitation to delocalize between the qubit $\{\ket{1}\}$, and the qudit $\{\ket{2},\ket{3}\}$. The interaction between them will contribute to the formation of LQU. This is strongly related to the system's quantum dynamics because, as the LQU accumulates, the transport efficiency is enhanced. Once the excess of noise inhibits the interaction between subsystems (due to localization), the state of the system becomes approximately pure and separable, and its contribution to LQU vanishes~\cite{Girolami}.}

\begin{figure}[H]
\centering
\includegraphics[scale=0.7]{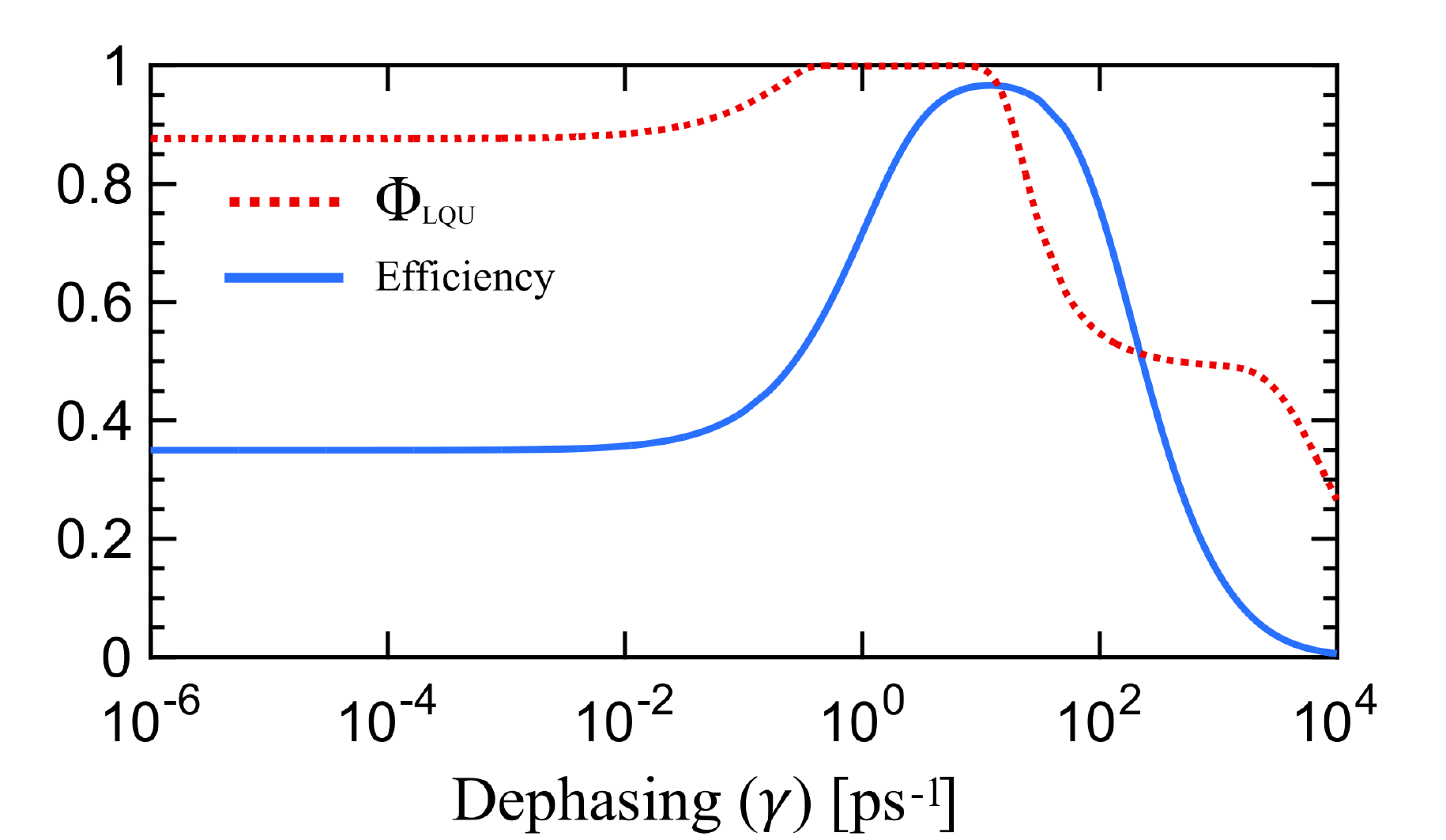}
\caption{Transport efficiency (blue solid line) and $\Phi_{LQU}$ (red dotted line) as a function of dephasing. Note that $\Phi_{LQU}$ reaches its maximum value in the same scale as the efficiency ($\gamma \sim$ 1--10), meaning that ENAQT and beyond-entanglement, discord-like quantum correlations are closely related.}\label{fig:figure2}
\end{figure}

To confirm that the LQU is strongly related to ENAQT in the FMO's subunit, we inspect, as a function of the dephasing rate, the interrelation between transport efficiency and LQU. Notice that as the efficiency $\eta$ is a cumulative value of the system, we compare it with the flux in LQU ($\Phi_{LQU}$) while $\gamma$ varies. 
Interestingly, Figure~\ref{fig:figure2} shows that the LQU flux follows a similar trend as the exciton-transport efficiency, i.e., it is enhanced by moderate dephasing and destroyed by stronger interactions with the environment. Notably, even though LQU is clearly related to ENAQT, not all the LQU resources are required to reach the maximum efficiency in the transport of excitations.

\textcolor{black}{Finally, given recent progress on noise-assisted transport in non-Markovian tight-binding networks~\cite{PRA_Ancheyta_2021}, we anticipate the presence of noise-assisted discord-like correlations in single-excitation systems undergoing non-Markovian dynamics. This is because non-Markovian environments have shown to increase either the quantum transport~\cite{MoreiraPRA2020} or the range of the dephasing rates where it occurs~\cite{PRA_Ancheyta_2021}.}

%

\section{Conclusions}
In summary, using a subsystem of the FMO complex, we have shown that beyond-entanglement quantum correlations emerge in the interaction of a photosynthetic system with its environment. Remarkably, by quantifying these correlations through a discord-like measure, the so-called Local Quantum Uncertainty, we found that these kinds of correlations follow the same trend as ENAQT, thus confirming the closed relationship that exists between them.  Our results may help to elucidate the role of discord-like quantum correlations in light-harvesting photosynthetic systems. Furthermore, they can be relevant in searching for quantum-enhanced applications relying on quantum information resources other than quantum entanglement.

\appendixtitles{no} 
\appendixstart
\appendix
\textcolor{black}{\section{}\label{app1}}
For the sake of completeness, in this appendix we show the derivation of the single-excitation LQU measure, obtained from the general definition presented in Equation (\ref{eq:eq7}). Note that, as Equation (\ref{eq:eq8}) dictates, the matrix elements of $W_{AB}$ are given by
\begin{equation}\label{eq:eqA1}
(W_{AB})_{ij}\equiv {\rm Tr}\{\sqrt{\rho_{_{AB}}}(\sigma_{iA}\otimes \mathds{1}_{B})\sqrt{\rho_{_{AB}}}(\sigma_{jA}\otimes \mathds{1}_{B})\}.
\end{equation}

To compute Equation (\ref{eq:eqA1}), we write the \textcolor{black}{square root of the} density matrix that comprises all \textcolor{black}{the $N$-}possible three-qubit states, i.e., $\rho_{_{AB}}$, in terms of its orthonormal eigenvectors $\ket{v^{AB}_{\ell}}$, with corresponding eigenvalues $\lambda_{\ell,AB}$, as
\begin{equation}\label{eq:eqn11}
\sqrt{\rho_{_{AB}}}=\sum_{\ell=1}^{N}\lambda_{\ell,AB}^{1/2}\ket{v^{AB}_{\ell}}\bra{v^{AB}_{\ell}}.
\end{equation}

If we now restrict the possible states of the system to the single-excitation basis, we can write its eigenvectors as
\begin{equation}\label{eq:eqA3}
\ket{v_{\ell}}= v_{\ell}^{1}\ket{100}+v_{\ell}^{2}\ket{010}+v_{\ell}^{3}\ket{001},
\end{equation}
where the coefficients $v_{\ell}^{n}$ (with $n=1,2,3$) are defined by the elements of the chromophoric system's density matrix, i.e., $\rho$. Note that the states in Equation (\ref{eq:eqA3}) are related to the states in Equation (\ref{eq:eq1}) by writing $\ket{1}=\ket{100}$, $\ket{2}=\ket{010}$ and $\ket{3}=\ket{001}$.  {Owing to the fact that} in the single-excitation basis the only non-vanishing element is $(W_{AB})_{zz}$, we can readily compute the LQU in {this basis} as
\begin{equation}
\mathcal{U}_{S}(t) = 1 - {\rm Tr}\left[\sum_{\ell,m=1}^{3}\lambda_{\ell}^{1/2}\lambda_{m}^{1/2}\ket{v_{\ell}}\bra{v_{\ell}}\sigma_{zA}\otimes\mathds{1}_{B})\ket{v_{m}}\bra{v_{m}}\sigma_{zA}\otimes\mathds{1}_{B} \right],
\end{equation}
with $\lambda_{\ell}$ being the eigenvalues of $\rho$.  {Since} the trace of an operator is the same irrespective of the basis in which it is expressed, we evaluate it in terms of the eigenbasis of $\rho$, yielding
\begin{adjustwidth}{-0.7cm}{0cm}\vspace{-6pt}
\begin{eqnarray}
\mathcal{U}_{S}(t) &=& 1 - \sum_{n=1}^{3}\left[\sum_{\ell,m=1}^{3}\lambda_{\ell}^{1/2}\lambda_{m}^{1/2}\braket{v_{n}|v_{\ell}}\bra{v_{\ell}}\sigma_{zA}\otimes\mathds{1}_{B})\ket{v_{m}}\bra{v_{m}}\sigma_{zA}\otimes\mathds{1}_{B}\ket{v_{n}}\right],
\nonumber\\
&=& 1-\sum_{l,m=1}^{3} \lambda_{l}^{1/2} \lambda_{m}^{1/2}\left| \braket{v_{l}|\sigma_{zA}\otimes\mathds{1}_{B}|v_{m}} \right| ^{2}.
\end{eqnarray}
\end{adjustwidth}
\end{paracol}

\reftitle{References}


\externalbibliography{yes}


\end{document}